 \documentclass[sigconf,authorversion]{acmart}

\def\R{\mathbb{R}}
\def\1{\mathbf{1}}

\def\E{\mathbb{E}}
\def\N{\mathbb{N}}

\usepackage[mathscr]{euscript}

\DeclareMathOperator*{\argmax}{arg\,max}

\newtheorem{definition}{Definition}

\usepackage{array}

\AtBeginDocument{%
  }

\copyrightyear{2025}
\acmYear{2025}
\setcopyright{acmlicensed}\acmConference[FAccT '25]{The 2025 ACM Conference on Fairness, Accountability, and Transparency}{June 23--26, 2025}{Athens, Greece}
\acmBooktitle{The 2025 ACM Conference on Fairness, Accountability, and Transparency (FAccT '25), June 23--26, 2025, Athens, Greece}
\acmDOI{10.1145/3715275.3732043}
\acmISBN{979-8-4007-1482-5/2025/06}

\begin{document}

\title[Push and Pull]{Push and Pull: A Framework for Measuring Attentional Agency on Digital Platforms}

\author{Zachary Wojtowicz}
\affiliation{%
  \institution{Harvard University}
  \city{Cambridge}
  \country{USA}}
\email{zwojtowicz@fas.harvard.edu}

\author{Shrey Jain}
\affiliation{%
  \institution{GETTING-Plurality Research Network}
  \city{Cambridge}
  \country{USA}}
\email{shreyjaineth@gmail.com}

\author{Nicholas Vincent}
\affiliation{%
  \institution{Simon Fraser University}
  \city{Burnaby}
  \country{Canada}}
\email{nvincent@sfu.ca}

\renewcommand{\shortauthors}{Wojtowicz \emph{et al.}}

\begin{abstract}
    We propose a framework for measuring \textit{attentional agency}, which we define as a user's ability to allocate attention according to their own desires, goals, and intentions on digital platforms that use statistical learning to prioritize informational content. Such platforms extend people's limited powers of attention by extrapolating their preferences to large collections of previously unconsidered informational objects. However, platforms typically also allow users to influence the attention of other users in various ways. We introduce a formal framework for measuring how much a given platform empowers each user to both \emph{pull} information into their own attention and \emph{push} information into the attention of others. We also use these definitions to clarify the implications of generative foundation models and other recent advances in AI for the structure and efficiency of digital platforms. We conclude with a set of possible strategies for better understanding and reshaping attentional agency online.
\end{abstract}





\maketitle

\section{Introduction}

\vspace{2ex}
\begin{quote}
        \emph{``The summation of human experience is being expanded at a prodigious rate, and the means we use for threading through the consequent maze to the momentarily important item is the same as was used in the days of square-rigged ships... Consider a future device... in which an individual stores all his books, records, and communications, and which is mechanized so that it may be consulted with exceeding speed and flexibility. It is an enlarged intimate supplement to his memory.''}

        \vspace{.5em}\hspace{4em}---Vannevar Bush, \emph{As We May Think}, 1945
\end{quote}

\vspace{2ex}

The original promise of the web was to create an interconnected network of information that could, in the words of Tim Berners-Lee, ``grow and evolve with the organization and the projects it describes'' \citep{berners1989information}. Early advocates of expanding the web ``world wide'' via the internet argued that limited access to information was a substantial bottleneck constraining both technological innovation and economic development. It seemed to follow logically that making the world's informational resources accessible online would greatly expand what humanity could accomplish.

In the decades since, the internet has grown precipitously, yet its originally envisioned benefits remain only partially realized. As it turns out, simply increasing the volume of freely available information online has not automatically produced a commensurately more well-informed global populace, in part because attentional constraints sharply limit what people can actually absorb and process  \cite{loewenstein2023economics}. As Herbert Simon \citeyearpar{simon1971designing} so presciently pointed out in the early days of the information age, decades before the world wide web was introduced:

\begin{quote}
    \emph{``In an information-rich world, the wealth of information means a dearth of something else: a scarcity of whatever it is that information consumes. What information consumes is rather obvious: it consumes the attention of its recipients. Hence a wealth of information creates a poverty of attention and the need to allocate that attention efficiently.''}
\end{quote}

Thus, attentional limitations mean that the practical utility of the internet depends critically on our methods for organizing and navigating the vast quantities of information it contains. Indeed, the limitations imposed by human attention mean that adding information to the internet without concomitant advances in search technology can actually reduce its overall functionality. Past a certain scale, attentional limitations necessitate curation, which, in turn, inevitably grants influence to whoever controls the curatorial process. We call this the internet's \emph{fundamental attention problem}.

In theory, the internet is a neutral suite of protocols that enable people to freely transmit information between computers on demand. In practice, however, a collection of platforms built on top of these core protocols mediate which information flows actually take place. A key insight motivating the present work is that many prominent platform categories (\emph{e.g.}, search engines, content libraries, social media sites, matching platforms, online marketplaces, and, increasingly, consumer AI systems) all create value by helping people overcome the internet's fundamental attention problem outlined above. 

In fact, many platforms accomplish this goal in essentially the same way---namely, by inferring people's attentional preferences, then algorithmically extrapolating them to intractably large sets of unconsidered informational objects (\emph{e.g.}, websites, products, people, content, messages, stories, and posts; see Table~\ref{tab:platforms}) \citep{resnick1997recommender,shapiro1999information}. To echo Vannevar Bush, platforms have become the ``means we use for threading through the consequent maze to the momentarily important item'' \cite{bush1945we}.


The algorithms that platforms employ to infer and extrapolate users' attentional preferences play a pivotal role in shaping who has agency on the internet and, increasingly, the world at large. In this paper, we introduce a formal framework for measuring the \emph{attentional agency} that a platform accords to each of its users. We define attentional agency as the capacity to allocate attention according to personal desires, goals, and intentions. Our framework offers a mathematical formulation of this concept that: (1) clarifies how various institutions currently solve the internet's fundamental attention problem, (2) furnishes operational measures of attentional agency that can be used to identify differences across users for any given platform, and (3) lends insight into how generative foundation models (GFMs) and other advances in artificial intelligence will impact the current paradigm for matching information and attention on the internet.

\subsection{Summary of our framework}

In using the term ``agency'' throughout this paper, we intentionally maintain a narrow focus on people's capacity to influence what they and others see on a given digital platform, acknowledging that this leaves out other potentially rich notions of agency (\emph{e.g.}, the specific actions a platform makes available to users).\footnote{We do not explicitly model feedback loops by which people's preferences are changed by their exposure to platforms and ranking systems. This is a critical area of future work, but as we will discuss below we believe even a relatively ``static'' analysis of attentional agency as described here provides immediate value.} The motivation for this approach is that it yields measures that are immediately calculable using information that platforms already gather in the process of delivering services to users, and can therefore be readily implemented by technologists and policymakers.\footnote{Our core approach and definitions are more general than this, however, as they can be applied to any utility function over informational objects for agents and advocates.}

The fact that many prominent platforms are oriented around solving the same underlying preference inference and extrapolation problem enables us to formalize attentional agency using a common mathematical framework---a generalization of the ``discounted cumulative gain'' measure used in ranking systems \cite{jarvelin2002cumulated}. The general nature of this approach enables us to abstract away from many idiosyncrasies that distinguish platforms (\emph{e.g.}, whether they sort websites, videos, or restaurants) and address a fundamental question: who has attentional agency? Specifically, our analysis begins with the observation that many information flows on the internet take the form of a stylized interaction between three parties: 

\begin{enumerate}
    \item An \emph{agent} seeking to allocate their limited attention to informational objects that they themselves value highly. Prototypical examples include a user searching the web, browsing a content library, or scrolling through a social media feed. 

    \item An \emph{advocate} seeking to influence what enters the agent's limited attentional field. Prototypical examples include an advertiser, government agency issuing a public-service announcement, or influencer.

    \item A \emph{platform}, which seeks to provide value to both agents (by inferring and extrapolating their preferences to large collections of mostly unconsidered informational objects) and advocates (by allowing them to influence what agents ultimately attend to). Prototypical examples include search engines, content libraries, online marketplaces, social media platforms, matching platforms, and, increasingly, consumer-focused generative AI products.
\end{enumerate}

\noindent On some platforms, agents and advocates are non-overlapping groups of users (\emph{e.g.}, job-seekers versus employers on a job board); on others, however, they are the same users acting in different capacities, either simultaneously or at different points in time (\emph{e.g.}, users of a dating app who both view profiles and want their own profile to be widely viewed).

As we show in greater detail below, when an agent and advocate have sharply conflicting priorities, a platform often cannot deliver maximal value to both simultaneously. An influencer trying to ``make a name for themselves,'' for example, might want their work to be seen by as many people as possible, while other users only want to see content that is related to their personal interests. We define two separate measures that capture how much attentional value a platform delivers to both agents and advocates. 

\begin{enumerate}
    \item[\textbf{Pull:}] The amount of attentional value a platform delivers to an agent as a fraction of what it is technologically prepared to deliver them given its current knowledge of agent preferences.
    
    \item[\textbf{Push:}] The amount of attentional value a platform delivers to an advocate as a fraction of what it is technologically prepared to deliver them given its current knowledge of agent preferences.
\end{enumerate}

Push and pull are not necessarily zero sum. Indeed, many forms of social interactions allow people to unilaterally enter into the attention of others (\emph{e.g.}, the practice of sending a cold-email, which can have net positive value for the recipient). In practice, however, increasing push often does reduce pull on digital platforms: limited attention means that pushing information into an agent's attention tends to exclude other objects they would have preferred to see more. As we elaborate in Section~\ref{sec:analysis}, the relative alignment of agent and advocate preferences and other primitives of a platform environment determine the rate at which these two measures trade off against one another.

Our framework defines pull and push at the level of an individual user for a given platform. These estimates can then easily be aggregated across users or systems. For instance, our push and pull measurements could be used to measure the attentional agency of specific sub-populations as defined by community membership, age, race, or other characteristics. Importantly, we define our measures such that the representations that platforms already compute in the process of mediating informational pushes and pulls can be used to quickly estimate different levels of agency---they are therefore not only operationalizable, but in a meaningful sense already operationalized.

Our framework highlights two separate channels by which GFMs and other advances in AI are impacting the nature of attentional agency online. First, platforms will ascertain progressively more granular representations of people's preferences and intentions. From the perspective of attentional agency, this is a double-edged sword---more accurate representations of a user's type not only enable platforms to deliver more targeted informational pulls, but also more targeted pushes. 

Second, GFMs excel at inferring the significance of complex digital objects at the level of their atomic informational components (``tokens'') and reconstituting them in novel ways \cite{bommasani2021opportunities,bubeck2023sparks,yin2023survey}. These models therefore enable platforms to extract and recombine information more flexibly. Whereas traditional search engines rank relevant web pages in their entirety, AI-powered search tools now recombine information from a variety of different pages into new informational objects on demand. This poses a fundamental challenge to the existing social, economic, and legal institutions that have emerged to generate and capture value on the internet. Most notably, the ability of GFMs to surgically extract information from its original delivery format---thus leaving embedded advertising, branding, and other promotional material behind---undermines many existing methods that platforms and content providers rely on to monetize their services. This is especially disruptive of the digital advertising model, which relies on the fact that people apprehend traditional informational bundles (\emph{e.g.}, web pages) as an integral whole. Stripping away context can also undermine established norms and expectations about how information should be interpreted, modified, and shared \cite{jain2023contextual}.


The rest of the paper is organized as follows. In the rest of Section 1, we provide a high-level summary of our framework, with additional motivation and an early coverage our limitations. In Section 2, we review related literature. Section 3 then presents our framework more formally and develops a mathematical analysis of our definitions. Section 4 highlights the broad applicability of our framework to five major categories of current internet platforms---search, content, social media, matching, and marketplaces. Section 5 presents some of the immediate policy implications of the concepts and formal definitions we propose and highlights open research questions.

\section{Related Work}\label{sec:related-work}

Our work builds on ideas from a variety of literatures in computer science, economics, and behavioral science. Below, we summarize key prior work, focusing first on proposals to better align digital platforms. Next, we discuss how our framework builds on prior literature evaluating the efficacy and fairness of ranking systems. Finally, we summarize work from psychology and economics on the analysis of individual preferences, which form a key element of our approach.

\subsection{Aligning Digital Platforms}

The challenge of aligning digital platforms has emerged as a key concern in modern computer science, prompting a variety of authors to propose methods for operationalizing potential social objectives, such as beneficence, equity, and fairness \citep{Kazim2021-ej,Shelby2023-qh}. The growing capabilities of GFMs have, moreover, introduced new forms of misalignment and new risks associated with failing to achieve alignment. A variety of potential solutions have been proposed in the literature, such as subjecting AI systems to a fiduciary requirement that would require them to both accurately learn people's preferences and uphold general principles, such as loyalty and care \cite{Benthall2023-cp}. Other related proposals include granting users new avenues of contestability \cite{Rakova2023-uu} and implementing fair ranking in domains like education \cite{Hafalir2021-rb,bar2023algorithmic}.

Conceptually, our article responds to a call originally made by \citeauthor{stray2021you} \shortcite{stray2021you}, who argue that recommender systems represent a specific instance of this broader value alignment problem. They specifically advocate for the creation of operational measures of alignment---developing such a measure is the central aim of this paper.

\subsection{Evaluation of Ranking Systems}

Our approach centers on a key feature of existing ranking and recommendation systems \cite{Herlocker2004-pi}, namely the user-specific ``relevance'' scores that platforms use to match user attention and informational content. Existing approaches to evaluating recommender systems include measuring the precision and recall of a list of a fixed size (\emph{e.g}., the ``ten blue links'' returned by a search engine), considering the \textit{mean reciprocal rank} (applicable to social media feeds of arbitrary extent), or calculating variants of the \textit{discounted cumulative gain}, which calculate a weighted average of relevance over ranks.

Our work intersects with prior research aimed at understanding how the design of recommender systems influences outcomes of interest, such as item diversity, bias, and user satisfaction  \cite{parapar2021towards,mcauley2022personalized}. Given that the business value of ranking is typically proprietary and somewhat difficult to measure, estimates vary widely \cite{Jannach2019-az}. Published work from Netflix suggests that the benefits of ranking-based personalization can be quite large in practice, at least in some domains \cite{Alvino2015-yy,Gomez-Uribe2016-lp}.

Our measure of attentional agency advances the literature on ranking evaluation by directly mapping this construct to relevance scores using a calculation that can be easily performed using existing infrastructure and institutional practices.

\subsection{Fair Ranking}

A large body of literature in machine learning has highlighted techniques to make ranking systems more ``fair,'' as interpreted in a variety of ways  \cite{dwork2012fairness,hardt2016equality,kleinberg2016inherent,diakopoulos2016accountability,wang2023survey}.

Patro \emph{et al.}\ \cite{patro_fair_2022} critically review a large body of work focused on fairness for ranking, recommendation, and retrieval systems. They highlight the inherent challenge of defining fairness in the face of a ``gap between high ranking placements and true provider utility, spillovers and compounding effects over time, induced strategic incentives, and the effect of statistical uncertainty'' (pg. 1929). Relatedly, Chen \emph{et al.}\ \cite{chen2023bias} provide a synthesis of how ``bias'' has been studied in the context of recommendation systems, arguing that group unfairness is but one of seven types of bias. 

Recent literature on fair ranking has underscored the difficulty of promoting fairness through a single definition, especially given that perception of fairness can vary widely from person to person \cite{alkhathlan_balancing_2024}. Techniques for incorporating fairness into ranking also tend to face a trade-off between fairness and utility (albeit techniques to ameliorate such trade-offs have also been developed \cite{dinh_learning_2024}). One goal of our proposal is to draw  the fairness community's attention to attention---that is, to highlight attentional agency as a potentially important factor in people's appraisals and experiences of fairness on digital platforms.

\subsection{Economic, Psychological, and Social Value of Attention}\label{sec:econ-psy}

People's preferences over what they pay attention to arise from a wide variety of emotional \citep{loewenstein2023economics}, cognitive \citep{tomasello2005understanding}, social \citep{braudy1997frenzy}, decisional \citep{stigler1961economics}, and other considerations. Our framework distills these diverse motivations into a single utility function and asks: to what degree do platforms empower people to satisfy their own attentional priorities? 

By operationalizing attentional agency in terms of inferred utility values, our framework inherits both the advantages and limitations of revealed preference analysis. We adopt this approach because it enables us to capture a wide variety of motivations in a single measure that can be inferred from people's behavior using standard techniques (indeed, the very techniques that platforms already use to deliver value to users).

Revealed preference measures do, however, leave open the question of how tightly coupled people's attentional preferences are to their personal well-being or broader notions of social welfare \cite{kleinberg2024inversion}. Indeed, there are many situations where the two seem to diverge quite sharply, such as video game addiction or ``doom scrolling'' highly engaging but emotionally deleterious content. Although our framework furnishes an operational measure of attentional agency, the question of how to evaluate its desirability in various circumstances is nuanced and requires careful consideration. Some of these concerns are only becoming more acute as AI search and other innovations enable platforms to deliver narrowly targeted persuasive messaging \citep{durmus2024persuasion,salvi2024conversational} and other novel forms of harmful content to users.

\section{Formal Framework}\label{sec:model}

\subsection{General Setup and Key Definitions}

We study a canonical interaction between three parties: an \emph{agent}, who wants to pay attention to informational objects they themselves value highly; an \emph{advocate}, who also has preferences about what the agent attends to; and a \emph{platform} that determines whether and how informational objects enter the agent's attention.

The \emph{agent} has type $\theta$ drawn from a distribution $F \in \Delta(\Theta)$ where $\Theta$ is a finite type space. There is a universe of informational objects represented by an ordered set $X = (x_1,x_2,\dots)$. An \emph{allocation} is a bijective function $\phi: \N \to X$. 

The agent has a type-dependent value for each informational object that is represented by a utility function $u:X \times \Theta \to \R_{\geq 0}$. Owing to their limited attention, the agent discounts informational objects based on how they are allocated according to a weakly decreasing function $\delta: \N \to [0,1]$, where $\delta_0 = 1$ without loss of generality. The agent's value of allocation $\phi$ is
\begin{equation}
    U(\phi, \theta) = \sum_{n \in \N} \delta_n \, u(\phi_n,\theta) 
\end{equation}

Logarithmic functions such as $\delta_n = \log_2(n+1)^{-1}$ have been studied extensively in the literature on ``discounted cumulative gain'' and are particularly influential in the design of ranking algorithms \cite{jarvelin2002cumulated}. More generally, different specifications of $\delta$ enable the model to capture the reduced-form effects of limited attention in a variety of problem domains and platform interfaces. For example, discounted cumulative gain is natural in situations where the agent sequentially attends to informational objects (\emph{e.g.}, a web search or social media feed); however, a ``cutoff'' function where $\delta_n = 1$ for $n \leq N$ and $\delta_n = 0$ otherwise might be more natural for an interface that simultaneously presented a handful of recommendations (\emph{e.g.}, a homepage featuring $N$ personalized suggestions, or an AI-augmented search result that synthesizes information into a block of text, all of which is presumed to be read).

The \emph{platform} supplies the agent with an allocation. Let $\Gamma$ denote a partition of $X$, which induces what we will call an \emph{allocation technology}.\footnote{A partition of $X$ is a set of subsets of $X$ (``partitions'' or ``blocks'') such that every element of $X$ belongs to one and only one block.} Specifically, we assume that the platform can produce allocations that rearrange the order of the partition blocks of $\Gamma$, but not move objects within blocks. The interpretation is that digital information is naturally bundled together into objects (\emph{e.g.}, web pages, academic articles, social media posts), and greater degrees of technological sophistication are required to break apart, comprehend, extrapolate preferences to, and recombine more granular units of information within these objects. Let $\Phi(\Gamma) = \{\phi:\N \to X \,|\, \forall \gamma \in \Gamma,\text{ if } x_i,x_j \in \gamma \text{ then } \phi^{-1}(x_i) - \phi^{-1}(x_j) = i - j\}$ denote the set of all possible allocations that can be constructed by permuting the blocks of $\Gamma$.

The platform has an imperfect understanding of $\theta$ and, therefore, the agent's preferences over allocations. Let $m$ denote the platform's \emph{model} of the agent, which is a reduced-form representation of: (1) all information the platform possesses that pertains to the agent's type (\emph{e.g.}, as collected from past interactions, present context, and their query); and (2) whatever algorithms and other technology the platform uses to transform this information into a prediction about their type. Some systems enable users to explicitly reveal aspects of their stable or transient preferences (\emph{e.g.}, by selecting interests when first creating one's profile or entering a search query), whereas other systems implicitly learn a user's type from patterns of engagement (\emph{e.g.}, dwell time or likes in a feed-based interface).

The \emph{advocate} also has preferences about what the agent sees, which we represent with a second utility function $v:X \times \Theta \to \R_{\geq 0}$. Paralleling our notation for the agent, the advocate receives value 
\begin{equation}
    V(\phi, \theta) = \sum_{n \in \N} \delta_n \, v(\phi_n,\theta) 
\end{equation}
Note that we are assuming the advocate discounts utility using the same positional function as the agent. This is natural, for example, if the discount function is taken to represent the amount of attention the agent pays to each position and both parties value informational objects based on how likely they are to be attended to. We assume the platform chooses an allocation that maximizes a linear combination of agent and advocate utility subject to technological feasibility.
\begin{equation}\label{eq:max}
    \phi_\lambda \in \argmax_{\phi \in \Phi(\Gamma)} \quad \E\big[\lambda U(\phi, \theta) + (1-\lambda) V(\phi,\theta) \,\big |\, m \big]
\end{equation}
for a given weight $\lambda \in [0,1]$.\footnote{In general, the parameter $\lambda$ is a choice variable under the control of the platforms. However, even if platforms can freely select $\lambda$ in principle, the full range of choices may not be feasible in practice. Platforms operate within a broader ``attention economy'' and consequently must compete with other activities for user engagement (most notably, the services provided by competitors). Market forces therefore constrain the minimum and maximum amount of $\lambda$ a platform can select without losing agent engagement or advocate buy-in entirely.}\footnote{An important exception occurs when platforms intentionally withhold attentional value to induce users to behave in a certain way—for example, when a dating app restricts how many profiles a user can view without paying for a premium version of the service. We note, however, that even in such cases, platforms typically incentivize behaviors by promising to move users onto or along the optimal push-pull frontier defined by the maximization \eqref{eq:max}.} For simplicity, we will assume throughout that the primitives of the model are such that the solution to this problem exists and is unique (or can be arbitrarily chosen from a set of candidate solutions; this assumption does not materially affect the points we raise in our discussion). Next, define
\begin{align}
    U_\lambda &= \E\big[ U(\phi_\lambda,\theta) \,\big|\, m \big] & 
    V_\lambda &= \E\big[ V(\phi_\lambda,\theta) \,\big|\, m \big]
\end{align}
which measure the amount of value that a particular choice of weight $\lambda$ accords to the agent and advocate, respectively. Note that, from the platform's perspective, the total expected value generated at a particular value of $\lambda$ is given by $P_\lambda = U_\lambda + V_\lambda$. 
We are now ready to define our central measures of attentional agency.

\begin{definition}
    \textbf{Pull} is the amount of informational value a platform delivers the agent as a fraction of the total value it is technologically capable of delivering them. Formally, for a given $u,v,\lambda$, $\Gamma$, and $m$, a platform's pull is
    \begin{equation}
        \textbf{Pull} = \frac{U_\lambda}{U_1}
    \end{equation}
\end{definition}

\begin{definition}
    \textbf{Push} is the amount of informational value a platform delivers the advocate as a fraction of the total value it is technologically capable of delivering to them. Formally, for a given $u,v,\lambda$, $\Gamma$, and $m$, a platform's push is
    \begin{equation}
        \textbf{Push} = \frac{V_\lambda}{V_0}
    \end{equation}
\end{definition}

\textbf{Push} and \textbf{Pull} both range from $0$ to $1$ and index the amount of attentional value that is being delivered to agents and advocates, respectively. According to these definitions, it is always the case that increasing push (weakly) decreases pull. Note, however, that push and pull are not necessarily zero-sum. We analyze the conditions that determine the relative tradeoffs between push and pull fully in the next section.

Another important point is that our framework defines push and pull for a single agent, advocate, and platform interaction. Across many interactions, each platform produces a distribution of $\textbf{Pull}$ and $\textbf{Push}$ values which could be analyzed in terms of their mean, variance, group-level differences, and other distributional properties. 

\subsection{Analysis}\label{sec:analysis}

In this section we highlight a few of the framework's most salient features. 

\subsubsection{Fixing technology, preference alignment determines the feasible push-pull frontier.}

First, note that $U_\lambda$ is a weakly decreasing function of $\lambda$ and $V_\lambda$ is a weakly increasing function of $\lambda$.\footnote{This is a direct consequence of the envelope theorem. See \cite{carter2001foundations}, pg. 603. for a formal statement and introduction to the theorem.} However, the rate at which these functions change depends upon the relative alignment of agent and advocate preferences. Taking the platform's allocation technology and agent model as given, we can consider the frontier of feasible push and pull values as a function of these underlying preferences. The platform effectively selects a point on this frontier by setting $\lambda$.

In a ``zero-sum'' situation where $u = -v$, the platform's objective simplifies to $(2\lambda - 1) \, \E[U(\phi,\theta)|m]$, and they must choose either to deliver the agent's most-preferred allocation (if $\lambda > \frac{1}{2}$) or their least-preferred allocation (if $\lambda < \frac{1}{2}$). In such an interaction, it is not only the case that pull and push are zero sum ($\textbf{Pull} + \textbf{Push} = 1$), but also that they change sharply around a critical value of $\lambda$. In other words, when agents and advocates push and pull in opposite directions, the platform will only accord agency to one of them. Conversely, if preferences are fully coincidental ($u = v$), then the platform can achieve maximal pull and push simultaneously ($\textbf{Pull} = \textbf{Push} = 1$) for any value of $\lambda$ because the same allocation is optimal for both agents.

A less intuitive case of simultaneously maximal pull and push occurs when the agent is indifferent over certain aspects of their allocation that the advocate has strict preferences over and vice-versa. When preferences are ``orthogonal'' in this way, the platform can also achieve maximal pull and push simultaneously ($\textbf{Pull} = \textbf{Push} = 1$) for any interior value of $\lambda$. From a policy perspective, situations of this variety are ideal targets for ``nudges'' \cite{thaler2009nudge}, given that they can, in theory, achieve a planner's objectives without decreasing people's attentional agency. 

In the more general case where preferences are partially aligned, the platform must choose how much weight they wish to place on agent versus advocate informational value, and, accordingly, how much agency each should be allocated. Note, however, that from the platform's perspective, the total informational value they create $P_\lambda$ is a concave function of $\lambda$.\footnote{This is another direct consequence of the envelope theorem.} This perhaps helps explain why interior values of $\textbf{Pull}$ and $\textbf{Push}$ are so commonly observed in practice---there is usually some value to the platform of choosing an interior $\lambda$ and dividing attentional agency to some extent.

\subsubsection{Model advances increase platform value, but also redistribute attentional agency}

Advances in machine learning and artificial intelligence (\emph{e.g.}, new architectures, more parameters, or additional compute \cite{kaplan2020scaling}) enter our framework in two ways. The first is through $m$, which represents the depth and specificity with which the platform infers an agent's type $\theta$ before constructing an informational allocation.
\footnote{The generality of this framework means that it can be applied to any recommender system, regardless of the specific algorithm used. See, \emph{e.g.}, \cite{Cai2022-na} for detailed treatment of how a seller might learn agents' types in a mechanism design context and \cite{mcauley2022personalized} for a general overview of user modeling.} 

Critically, however, such changes enhance the platform's ability to provide value to both agent and advocate. In the former case, accurate inferences about an agent's type enable the platform to more successfully extrapolate their preferences to hypothetical allocations. Simultaneously, these inferences also enable the platform to facilitate more precise targeting by advocates. Indeed, recent work shows that frontier models are capable of greater persuasion precisely because they are better at ``matching the language or content of a message to the psychological profile of its recipient'' \cite{Matz2024-nm}, increasing the stakes of strategic misalignment between agents and advocates.

The net effect of model improvements on attentional agency therefore depends on whether the innovations are more useful for predicting marginal differences in $u$ or $v$ across various allocations. Recall that the platform's objective when selecting an allocation is to maximize the linear combination
\begin{equation}
    \lambda \, \E[U(\phi,\theta) \,|\, m] + (1-\lambda) \, \E[V(\phi,\theta) \,|\, m]
\end{equation}
for some value of $\lambda$. If two types $\theta$ and $\theta'$ have conflicting preferences over allocations, then learning to distinguish them will make it more clear, in each case, what the marginal value of choosing one allocation over the other will be. This, in turn, will tend to shift emphasis in the overall optimization toward the agent's preferences. A parallel argument, however, is also true for targeting.

Although the net effect on pull and push will depend, in each case, on which of these effects prevails, one implication always holds: the \emph{stakes} of informational agency consistently grow as models become more sophisticated.

\subsubsection{AI enables attentional allocation to operate over more granular and varied information objects}\label{sec:analysis-granularity}

Advances in artificial intelligence also enter our framework through the partition $\Gamma$, which defines the units over which a platform can deliver information. 

Multiple features of modern generative foundational models \cite{bommasani2021opportunities} now enable AI systems to achieve unprecedented levels of informational granularity: (1) attentional transformers parse a wide variety of data sources into their constitutive tokens, meaning that systems can apprehend complex informational objects at the scale of individual symbols \cite {vaswani2017attention}; (2) expanded context windows enable sequence-based models to capture long-range data dependencies \cite{press2021train,su2024roformer}; (3) multi-modality is in the process of extending the informational purview of AI systems to text, image, video, audio, and other formats \cite{yin2023survey}; and (4) generative capabilities enable systems to flexibly recombine these informational atoms into arbitrary new structures that do not necessarily resemble the sources they were initially drawn from. Although it has always been possible for platforms to disassemble informational objects in a literal sense, foundation models enable a categorically greater degree of sophistication when extrapolating an agent's preferences and intentions to these smaller units of information.
\footnote{There may be cases in which an agent's utility function $u$ contains a preference for less granular partitions. For instance, consider an agent that prefers to listen to music \textit{only} in the form of complete and correctly ordered albums.}

Within our framework, the cumulative effect of these innovations is to refine the platform's informational domain into a finer partition $\Gamma' \subset \Gamma$. \footnote{Specifically, if for every $\gamma' \in \Gamma'$ there exists a $\gamma \in \Gamma$ such that $\gamma' \subseteq \gamma$ in the sense of weak subsetting, where at least one inclusion is strict.} The additional degrees of freedom this opens up expand the space of possible allocations and, consequently, enable the platform to deliver more value overall. Formally, $\Phi(\Gamma) \subset \Phi(\Gamma')$ so that ${P'}_\lambda \geq P_\lambda$ for any given $\lambda$. As in the case of model development, the ultimate impact of such a change on welfare and informational agency will depend on the relative alignment of preferences and other primitives of the model. 

One clear take-away, however, is that highly capable multi-modal AI models which are fully aligned to agents (\emph{i.e.}, for which $\lambda=1$ so that $\textbf{Pull}=1$) increase their ability to extract informational value from a variety of sources without satisfying the implicit ``attentional bargain'' that supports much of the internet's current economic model---branding, sponsored content, advertising, and other features that businesses employ to monetize their content. As \citep{wu2017attention} points out in his history of the attention economy, the basic strategy to ``draw attention with apparently free stuff and then resell it'' has long been a part of the advertising business. Technological innovations intermittently disrupt the \emph{status quo} implementation of this strategy, as exemplified by the famous ``Betamax case'' over whether Videocassette recordings of television shows (which enabled people to re-watch them without embedded advertisements) constituted copyright infringement.\footnote{\emph{Sony Corp. of America v. Universal City Studios, Inc.}, 1984.} 

In contrast to other drivers of creative destruction, however, the very generality of general artificial intelligence means that a single technology stands to disrupt nearly every existing informational format simultaneously. The advent of human or super-human level artificial intelligence may represent a final frontier in these debates, effectively solving the attention problem. We return to the question of how these insights inform policy approaches to refactoring the landscape of attentional agency in Section~\ref{sec:policy}, below.

\section{Attentional Agency on Platforms}\label{sec:platforms}

\renewcommand{\arraystretch}{1.75}
\setlength{\tabcolsep}{10pt} 
\begin{table*}[t]
    \centering
    \footnotesize
    \begin{tabular}{|
    >{\raggedright\arraybackslash}p{1.3cm}|
    >{\raggedright\arraybackslash}p{1.7cm}|
    >{\raggedright\arraybackslash}p{1.7cm}|
    >{\raggedright\arraybackslash}p{1.7cm}|
    >{\raggedright\arraybackslash}p{1.7cm}|
    >{\raggedright\arraybackslash}p{1.7cm}|}
    \hline
        & \textbf{Search} & \textbf{Content} & \textbf{Social} & \textbf{Matching} & \textbf{Marketplace} \\
    \hline
        \textbf{Attentional Act} & Enter search query & Browse library & Refresh feed & View potential matches & Enter product query \\
    \hline
        \textbf{Advocates} & Advertisers, SEO & Advertisers, platforms & Influencers, advertisers & Promoted users & Sellers, advertisers \\
    \hline
        \textbf{Type} \newline ($\theta$) & Demographics, location, context & Preferences over genres, formats, artists & Social interests, browsing patterns & Desired match, dating goals & Shopping needs, market trends \\
    \hline
        \textbf{Informational Objects} \newline ($X$, $\Gamma$) & Indexed websites, documents & Books, music, movies, podcasts, shows & Posts, stories, short-form video & Dating profiles, job listings & Product listings \\
    \hline
        \textbf{Model \newline ($m$)} & Query, search history, IP address, cookies & Viewing/listening history, ratings & Behavioral analytics, engagement & Preference matching & Past purchases, viewing history, dynamic pricing \\
    \hline
    \end{tabular}
    \caption{How prominent digital platform categories map onto our framework.}
    \Description{How prominent digital platform categories map onto our framework.}
    \label{tab:platforms}
\end{table*}

Table~\ref{tab:platforms} provides a high-level overview of how platform categories map on to our framework. Comparing various platforms side-by-side highlights not only their common approach to  attentional allocation, but also that the boundaries separating these categories are becoming increasingly blurred over time. Search engines that may have once depended purely on a user's entered query and a few contextual variables (\emph{e.g}. location) now incorporate user history \cite{Matthijs2011-wz}. Social media platforms often emphasize a feed (with minimal interaction options beyond ``show me more content'') but may also support search. Online marketplaces sell a mixture of physical and digital goods, but now often offer recommendation features.

In what follows, we briefly review how our framework maps onto existing platform categories, then briefly discuss how generative foundation models are further eroding the distinctions that have traditionally separated them \cite{Deldjoo2024-yf}.

\hspace{1.25em}\textbf{Web Search Engine}: Search engines primarily rank web pages based on user queries and intrinsic, site-specific features \citep{varian2016economics,resnick1997recommender}. Agents engaged in web search generally want to specify their query as briefly as possible, meaning that search engines are most efficient when they can anticipate people's needs and desires \citep{resnick1997recommender,Matthijs2011-wz}. This tends to incentivize platforms to build up a deep understanding of preferences and intention through repeated interaction with each individual user. One user's interactions with a platform are also used to tailor results delivered to similar users through techniques such as collaborative filtering.

Queries enable users to locate relevant content efficiently, but they also equip platforms with highly specific data about user characteristics and momentary intentions. Consequently, many leading search engines harness this information by allowing web pages to augment their ranking preferentially among users with specific characteristics.

\vspace{.5em}\textbf{Content Library}: Content libraries provide access to large collections of media and generally enable users to both search for items they already know and browse recommendations to discover new items. Media preferences tend to correlate significantly across and within individuals, meaning that both general popularity and personalization can significantly enhance the value of recommendations  \citep{gomez2015netflix}.

In the context of content libraries, pull could be decreased because of platform promotion (\emph{e.g.}, a video streaming service promoting a show they funded), but also through limitations of collaborative filtering.

\vspace{.5em}\textbf{Social Media}: Social media platforms present users with a curated stream of multimedia objects centered around the activities of other users. Many social media feeds orient the user's attention around a single activity: requesting more content. 
Social media platforms tend to primarily infer an agent's interests and intentions implicitly (\emph{e.g.}, from dwell time), rather than explicitly (\emph{e.g}, entering a query). Social media is also distinguished by the fact that many users want to both ``see and be seen,'' acting as both agents and advocates.

\vspace{.5em}\textbf{Matching}: Matching platforms help interested parties find one another. Some such platforms, such as dating apps, are ``two-sided,'' with users acting as both agents and advocates. Many matching platforms allow advocates (premium users) to gain visibility by influencing the attention of other users.

\vspace{.5em}\textbf{Marketplace }: Online marketplaces help people search for goods and services. User preferences on marketplaces are driven by a combination of immediate shopping needs, long-term consumption habits, and sensitivity to factors like price, quality, and seller reputation.
Attention can have a sizable impact on consumer behavior \citep{gabaix2006shrouded}, which makes featured listing, personalized recommendations, promotional offers, and targeted advertising especially valuable to advocates. 

\textbf{AI Search:} Table \ref{tab:gen_foundation_models} summarizes how AI search tools (powered by generative foundation models) compare to the five existing platform categories just reviewed. 
As discussed in Section \ref{sec:model}, AI search tools operate at the level of individual tokens \cite{bommasani2021opportunities}, which enables them to flexibly tailor content to the user's specific preferences ($\theta$), as inferred from their interaction history and context. The capacity of modern AI systems to draw meaningful connections between distal information sources and modalities (see Section~\ref{sec:analysis-granularity}) strengthens complementarities between various data sources:  scientific papers, law articles, encyclopedia entries, code, and other sources of knowledge that were previously mostly referenced independently can now be interrelated in new and more valuable ways.

\begin{table*}[htbp]
    \centering
    \small
    \begin{tabular}{
    |>{\raggedright\arraybackslash}p{4cm}|
    >{\raggedright\arraybackslash}p{8.5cm}|}
    \hline
        & \textbf{AI Search / Chat / Assistant} \\
    \hline
        \textbf{Attentional Act} & Request for information via a natural language prompt \\
    \hline
        \textbf{Advocates} & (As of yet undetermined on most platforms)\\
    \hline
        \textbf{Type ($\theta$)} & User preferences, needs, and intentions \\
    \hline
        \textbf{Informational Domain} ($X$, $\Gamma)$ & Sub-components of any other informational domain \\ 
    \hline
        \textbf{Model} ($m$) & Generative foundation model, context window, specific prompt \\
    \hline
    \end{tabular}
    \caption{Applying the push/pull framework to AI search, chat, and assistant tools, as powered by modern generative foundation models.}
    \Description{Applying the push/pull framework to AI search, chat, and assistant tools, as powered by modern generative foundation models.}
    \label{tab:gen_foundation_models}
\end{table*}

\section{Discussion}\label{sec:discussion}

\subsection{Advertising in the Technology Industry}

A common refrain in early commentary on the targeted digital advertising model was ``if you are not paying for the product, you are the product.'' Subsequent proposals, such as data as labor \cite{posner2018radical} and data dignity \cite{lanier2018blueprint}, pursue a related critique: that under the current paradigm, users of free internet platforms are also effectively workers producing data. Advocates of the advertising model, on the other hand, argue that it promotes equity and access because free products extend the benefits of technological innovation to those at the bottom of the socio-economic ladder.

Our framework shows how key claims frequently made in these debates can be made more concrete by quantifying the amount of attentional agency these platforms accord to various groups of individuals. If someone is not paying for a costly service, their activity is being subsidized in some way, often by advocates. The degree to which this reduces their attentional agency, however, depends on the relative alignment of preferences, technological context, and other factors. Measures of push and pull show how the combined influence of these factors can be operationalized using relatively lightweight calculations.

This creates an open challenge for balancing the welfare implications of attentional agency and the consumer surplus afforded by ad-funded internet platforms \cite{Brynjolfsson2023-py}. Below, we outline ideas aimed at making tangible progress on this challenge, especially in light of recent AI progress.

\subsection{Attentional Agency and Alignment}\label{sec:discussion-alignment}

Our measures of attentional agency also complement ongoing research into AI alignment, especially work that applies concepts such as pluralism in human values \cite{sorensen_value_2024} and social choice \cite{conitzer2024social}. Platforms create value by extending a user's attention to unseen informational objects, which necessarily entails the user entrusting the platform with a certain degree of discretion. Thus, pull can measure alignment to an individual's intentions. Push, on the other hand, measures alignment to the objectives of others.

Push and pull further clarify interventions that aim to block specific concepts or pieces of information (\emph{e.g.}, instructions to build a weapon), such as preference ranking optimization \cite{song_preference_2024} and chained ``safety classifiers'' \cite{kim2023robust}. In our framework, such measures can be understood as seeking to fulfill the goals of an advocate---which might be a public body acting in the public interest. This insight is useful because modeling such situations as conflicts between advocates allows us to bring economic reasoning and mechanism design to bear on the problem.

\subsection{Attentional Agency and Generative Foundation Models}\label{sec:policy-gfm}

Our framework reveals that, from the perspective of attentional agency, the key innovation of generative foundation models lies in their ability to parse information from a wide range of sources at an exceptionally granular level---that of the individual token. As we argued in Section~\ref{sec:analysis-granularity}, this enables AI-powered search tools to extract information from websites while leaving behind embedded advertising and other elements that support monetization.  This raises a critical question: will some form of the advertising model endure, or will it give way to a new paradigm in which agents pay directly (\emph{e.g.}, via subscriptions or API credits) for access to information?

The potential for generative foundation models to circumvent existing monetization pathways has raised concerns about their implications for journalism and related industries \cite{Grynbaum2023-ma}. Our framework highlights, however, that this problem arises precisely because these technologies enable agents to maximize pull. However, there exists a hopeful future in which direct access to highly personalized and carefully ranked chunks of information is worth paying for and ultimately leads to \textit{more} funding for information creation and curation. 

In short, if we view advances in AI as reducing the granularity of informational objects that can be processed on digital platforms, the lens of attentional agency gives us a clear vocabulary (and with regulatory support, concrete data---see below for how lightweight regulatory interventions could allow for the sharing and use of push and pull measurements) to reason about when the reduction in granularity will work to advance human values, and when it will create new challenges (\emph{e.g.}, violate previous arrangements around incentives).

\subsection{Policy Strategies}\label{sec:policy}

There have been many proposals to promote greater transparency and accountability online. Some of the most prominent examples include the transparency provisions in the EU's Digital Services Act (DSA), EU's General Data Protection Regulation (GDPR), and proposed EU Artificial Intelligence Act \cite{digital_services_act,gdpr} \footnote{Reporting of metrics that we outline in this paper are not new. For instance, Article 26 of the DSA requires platforms to provide users with meaningful information about the main parameters used in targeted advertising and ways to modify these settings. Article 26 also calls for a ban on targeted advertising based on profiling using special categories of personal data identified in Article 9 of the GDPR. Additionally, Article 27 of the DSA requires platforms to clearly explain the main parameters that influence what information is suggested to users and provide options to modify these parameters. Furthermore, Article 28 requires the largest online platforms and search engines to provide users with at least one recommender system option that is not based on profiling of individuals by automated systems as defined in the GDPR \cite{TechPolicyPress2022}}.

Below, we highlight three policy strategies that naturally complement our framework. Notably, each of these simple proposals can be implemented with relatively little effort or tailoring by platforms.

\begin{enumerate}
    \item \textbf{Transparency into attentional agency}: Measure and provide information on the distribution of $\textbf{Pull}$ and $\textbf{Push}$, both in aggregate and by subgroup. Greater transparency would help users identify instances where they are experiencing reduced levels of attentional agency and make informed choices between platforms, potentially increasing both accountability and competition.
    \item \textbf{Expanded consumer choice:} A variety of platforms sell push in order to generate revenue. For those that do, users could be given the option to pay for a ``push-free'' experience (which would generalize the ``ad-free'' tier offered by some services). This would enable users who place a greater premium on pull to maximize it for a price, increasing overall economic efficiency.
    \item \textbf{Provably neutral algorithms:} Platforms offer a version of their service independent of advocate-driven factors, but that may not maximize pull. An example would be the inclusion of an option to sort one's social media feed chronologically and without promotion.
\end{enumerate}

These strategies all target the concrete attentional choices that platforms make on behalf of users and the structural incentives that shape those choices. These ideas are not a substitute for existing algorithmic audits, privacy protections, content moderation standards, or other key planks of the online safety agenda. Attentional agency attacks the problem of manipulative platform design from a neglected angle, one grounded in a numerical accounting of how platform incentives distort the information that users encounter.

While paid push-free options, neutral algorithm defaults, or push/pull disclosures may seem like relatively modest interventions, they reflect a fundamentally different theory of platforms than conventional transparency approaches. Rather than focusing on the black-box mechanics of ranking and recommendation systems, they target the concrete attentional choices that platforms make on behalf of users and the structural incentives that shape those choices. Importantly, measures to increase transparency at this level would support broader societal conversations about how attentional agency is distributed without requiring detailed technical knowledge of \emph{how} algorithms produce specific rankings.

As we pointed out in our discussion of the ``Economic, Psychological, and Social Value of Attention'' (Section~\ref{sec:econ-psy}), people's revealed attentional preferences are not always tightly coupled to their underlying welfare. Policies that uncritically seek to maximize pull may therefore not only fail to improve, but in fact actively undermine, a user's overall health and happiness (\emph{e.g.}, if they are struggling with attentional self-control problems). One promising solution would be to develop and apply our framework to algorithms that infer people's reflective (rather than impulsive) preferences (see \citep{kleinberg2024inversion} for a general discussion).

\subsection{The Political Economy of Social Media and AI}

Recent policy debates have highlighted a variety of issues that arise when large organizations have the capacity to determine what people pay attention to at scale. As one example, concerns about the political independence of TikTok's content and algorithmic recommendations spurred the United States to pass legislation banning any ``website, desktop application, mobile application, or augmented or immersive technology application'' that is ``controlled by a foreign adversary'' \cite{uscongress2024}.

Our framework helps inform these debates by providing a quantitative tool that can be used to evaluate the impact that competing incentives of various stakeholders have on user agency. As just discussed in Section~\ref{sec:policy-gfm}, generative foundation models and other innovations are quickly expanding the capacities of platforms, raising the stakes of developing a more participatory role for users in technological development \cite{allen2024real}.

These developments reflect a growing desire for better evidence-based approaches to platform accountability and governance.  By demanding transparency into the attentional agency of users, policymakers and the public can work towards aligning platform incentives with democratic values and the genuine preferences of users.

Measures of attentional agency also complement recent work on the interplay of algorithmic platforms and phenomena such as ideological segregation \cite{gonzalez2023asymmetric}, echo chambers and ``rabbit holes'' \cite{brown2022echo}, and interaction with misinformation more generally \cite{edelson2021understanding}. This body of work provides a complementary lens for understanding the health of an algorithmic platform or feed. Given that push and pull measurements can be assessed in a faceted manner, it could be useful to consider push and pull along axes such as political valence or through the lens of ``trust.'' 

\section{Conclusion}

The framework we propose provides a lens for reasoning about how current platforms distribute attention. Our methods also provide new insight into ways that future platforms, especially ones powered by highly capable forms of artificial intelligence, might come closer to achieving the original visions of the thinkers who inspired the internet. The framework we propose is intended to be readily applicable to policy discussions---indeed, the operational definitions we propose are based on techniques and computations already in use by platform operators.

\bibliographystyle{ACM-Reference-Format}
\bibliography{bib}

\end{document}